\documentclass[doublecol]{epl2} 
\usepackage{graphicx}
\usepackage{bm} 
\usepackage{amsmath}
\usepackage{amssymb}
\usepackage{color}

\def\JB#1{{{#1}}}

\title{Localization by entanglement}

\author{Joachim Brand\inst{1} \and Sergej Flach\inst{2} \and Victor
  Fleurov\inst{2,3} \and 
  L. S. Schulman\inst{2,4} \and Denis Tolkunov\inst{4}}
\shortauthor{J. Brand\etal}

\institute{ \inst{1} Centre of Theoretical Chemistry and Physics,
  Institute of
  Fundamental Sciences, Massey University Auckland, New Zealand\\
  \inst{2} Max-Planck-Institut f\"ur Physik komplexer Systeme,
  N\"othnitzer Str. 38, D-01187 Dresden, Germany\\
  \inst{3} {School of Physics and Astronomy, Tel Aviv University,
    Israel}\\
  \inst{4} {Department of Physics, Clarkson University, Potsdam NY,
    USA} } 


\abstract{
We study the localization of 
bosonic atoms in an optical
lattice, which interact in a spatially confined region.
The classical theory predicts that there is no localization below a
threshold value for the strength of interaction that
is inversely proportional to the number of participating atoms.
In a full quantum treatment, however, we find that
localized states exist for arbitrarily weak attractive or repulsive
interactions for any number ($>1$) of atoms. We further show, using
an explicit solution of the two-particle bound state and an
appropriate measure of entanglement, that the entanglement tends to
a finite value in the limit of weak interactions. Coupled with the
non-existence of localization in an optimized quantum product state,
we conclude that the localization exists by virtue of entanglement.}

\begin{document}

\maketitle

Spatial localization of quantum interacting particles and formation of
bound states are of fundamental interest to modern physics. One
intriguing aspect is the correspondence between localized states in
classical and quantum mechanical theories \cite{zahed86:_skyrm}.
Usually, one expects quantum fluctuations to weaken localization, as
the binding of particles with an attractive but shallow pair potential
can be inhibited by quantum mechanical zero point motion.  Then,
localization can be interpreted essentially as a classical property
that would emerge in a quantum system due to decoherence
\cite{dunningham05:PedCatsFlufBun}. On the other hand, it was recently
suggested that localization of quantum particles may be achieved when
they are entangled through suitable measurements
\cite{rauh03:_locEntangl}. Here, {we consider the role of
  entanglement in the localization of specific eigenstates of a
  multiple boson system, e.g. the ground state. Specifically,} we show
that spatially confined interaction between atoms in an optical
lattice induces entanglement and leads to localization, while the
corresponding classical atomic field fails to localize. Remarkably,
the effect that we demonstrate in this Letter does not depend on
whether the interaction is attractive or repulsive. Recent experiments
on formation of repulsive atomic pairs on optical lattices
\cite{winkler06:_repul} imply the possibility of experimental
observation of the effect reported here.

If a translationally invariant lattice with interactions is
considered, its classical limit allows for localized solutions known
as lattice solitons or discrete breathers \cite{campbell04:_DB}. A
particular realization of such a system is a Bose-Einstein condensate
(BEC) in an optical lattice \cite{morsch06rmp}. Due to the band
structure with Bragg reflection gaps in the optical lattice, localized
soliton solutions are possible not only with attractive but also with
repulsive interactions. Experimental evidence for the band gap
solitons with repulsive BECs has been reported for one-dimensional
lattices \cite{eiermann04}. For two- and three-dimensional lattices
the classical theory predicts nonzero energy and particle number
thresholds for the existence of band gap solitons
\cite{flach97:thresholds}, as opposed to the case of dimension one.
Quantum effects in this system are expected to be most dramatic for a
small number of particles 
\cite{scott94:_quant,fleurov03chaos}.  The extreme quantum limit of a
three-dimensional lattice has been realized in the experiment of
Winkler {\it et al.}  \cite{winkler06:_repul}, where bound pairs of
repulsively interacting atoms have been reported using spectroscopic
tools.  According to quantum theory, these quantum solitons describe
bound states of atoms that delocalize spatially
\cite{lai89praI,lai89praII,chiao91prl-two-phot}. It is an open
question, whether these observed bound states persist below the
above-mentioned classical threshold.

In this Letter we study
localization of atoms in
{an}
optical lattice, where interactions between atoms are present in a
spatially confined region only. This can be achieved experimentally
by tuning the $s$-wave scattering length by the Feshbach resonance
with inhomogeneous magnetic \cite{Tiesinga93pra,Inouye1998a} or
laser fields \cite{Fedichev1996a,theis04prl:OpticalFR}. We show that
localization occurs in the full quantum system when it is forbidden
classically. The crucial difference between the quantum and the
classical models is the presence or absence of entanglement (see
\cite{popescu06:_entan, vedral06entanglement, paskauskas01pra})
between the constituent particles. The predicted quantum
localization is due to entanglement. For the 
case of two
particles we quantify the entanglement and show that it reaches a
finite value in the limit of weak interactions.
{Beyond the specific 
model studied we also comment on the relation between entanglement in 
eigenstates, localization, and the existence of bound states for
higher dimensional and translationally invariant systems.}
 It is worth
mentioning here that 
bound states for two electrons (fermions) in
the negative hydrogen ion also appear only 
in the presence of quantum correlations
beyond the
Hartree-Fock approximation\cite{hill77HIon}.

{\em The model} - We {initially} consider the dynamics of
atoms in 
{a one-dimensional}
optical lattice in which the atoms interact in a spatially confined
region. The Hamiltonian is given by
\begin{equation} \label{eqn:hamiltonian}
H = -\sum_n (a^\dagger_n a_{n+1} + a^\dagger_{n+1} a_{n}) + \lambda
a_0^\dagger a_0^\dagger a_0 a_0 ,
\end{equation}
where $a^\dagger_n$ ($a_n$) creates (destroys) a boson on the
lattice site $n$ and bosonic commutation relations $[a_n,
a_m^\dagger] = \delta_{nm}$ hold. The Hamiltonian
(\ref{eqn:hamiltonian}) describes bosonic atoms on a lattice that
interact either repulsively ($\lambda >0$) or attractively
($\lambda<0$) only on the single lattice site $n=0$. Single atoms
with the Hamiltonian (\ref{eqn:hamiltonian}) on a lattice with $M$
sites 
and periodic boundary conditions
do not localize and the eigenstates are plane waves
$1/\sqrt{M}\sum_n\exp(ikn)\; a_n^\dagger |{\rm vac}\rangle$ where
$|{\rm vac}\rangle$ is the vacuum state (no particles). However,
with more than one particle localized states may exist around the
site $n=0$.
In this current model the binding of particles implies spatial
localization and vice versa.

{\em Classical {treatment}
} -
The quantum Hamiltonian (\ref{eqn:hamiltonian}) can also be
understood as posing a {\em classical} Hamiltonian lattice problem
if we replace the particle creation and destruction operators by
complex valued functions of time. 
{In order to enable a detailed comparison between classical and
quantum predictions it is necessary to establish the precise relation
between both pictures.}
An unambiguous route to relate the classical with the quantum problem
can be found by the Hartree ansatz and variational procedure:
For
the many-body wave function with $N$ particles we use the ansatz of
a product state $|\Psi^{(N)}_{\rm H}\rangle = 1/\sqrt{N!}\,
(b^\dagger)^N|{\rm vac}\rangle$ where $b^\dagger = 1/\sqrt{N}\sum_n
\psi_n^* a_n^\dagger$ creates a single particle with the complex
amplitude $\psi_n$ on the lattice site $n$. The corresponding equation
emerges from the standard Lagrangian variational procedure with $g =
2 \lambda (N-1)/N$ assuming normalized solutions with $\sum_n
|\psi_n|^2 = N$. This equation is the discrete non-linear
Schr\"odinger (DNLS) model with nonlinearity present only on the
site $n=0$,
\begin{equation} \label{eqn:nls}
i \frac{\partial}{\partial t} \psi_n = -(\psi_{n+1} + \psi_{n-1}) +
g \delta_{0,n} |\psi_0|^2 \psi_n\;.
\end{equation}
This model was originally introduced to study the transport of
electrons coupled to lattice phonons \cite{Tsironis94pre}. The model
also applies to BECs in an optical lattice and has been discussed in
connection with Fano resonances in the transport of cold atoms
\cite{vicencio:184102}. Here, $\psi_n$ describes the complex matter
wave field at the lattice site $n$ after the introduction of
appropriately rescaled dimensionless variables. $N = \sum_n
|\psi_n|^2$ is the number of atoms in the BEC.

\JB{ We have now used the Hartree procedure to {\em derive} the set of
  classical equations (\ref{eqn:nls}) from the quantum problem
  (\ref{eqn:hamiltonian}). The same set of equations (\ref{eqn:nls})
  would have also emerged from a more standard approach using a
  coherent-state ansatz for the many-body wave function. However, the
  Hartree procedure here serves a dual purpose in also characterizing the
  classical equations (\ref{eqn:nls}) as an {\em approximation} to the
  quantum problem that provides strict variational bounds for the
  latter.}

\begin{figure}[hbt]
\center
\includegraphics[width=\columnwidth]{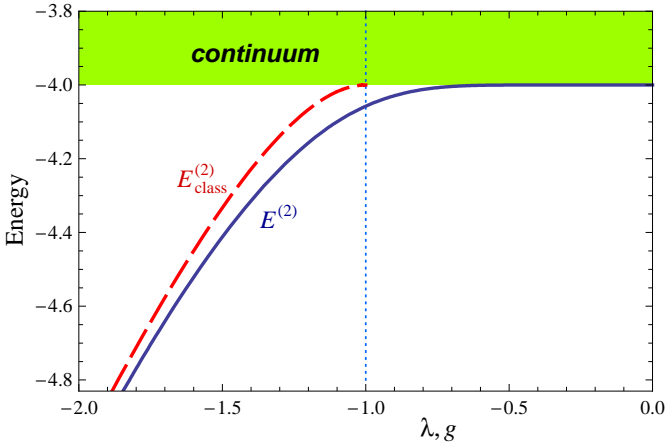}
\caption{Relation between energy and coupling
constant for two-particle defect states. The dashed line shows
the classical (Hartree) solution $E^{(2)}_{\rm class} = 2(g +
\frac{1}{g})$. The shaded  region at $E > -4$ indicates the
edge of the continuum band of linear waves. At the classical threshold
of $g = 1$ (the dotted
line) the classical solution reached the continuum edge. The
solid line shows the exact solution $E^{(2)}$ of the
two-particle problem of eq.~(\ref{eqn:Eexact}), which persist even
below the classical threshold down to zero coupling. }
\label{fig:FigThresh}
\end{figure}

The model of eq.~(\ref{eqn:nls}) supports plane wave solutions in
the linear ($g=0$) case
\begin{equation}
\psi_n^{\rm (pw)} = \psi_0 \exp(ikn) \exp(-i\omega t)
\end{equation}
with the dispersion relation $\omega = -2\cos k$ defining a band
continuum $[-2,2]$. In addition, for non-zero $g$, there are
localized solutions
\begin{equation}\label{eq:psiloc}
\psi_n^{\rm (loc)} = A e^{-\delta |n|} e^{-i \Omega t} e^{i \theta n},
\end{equation}
with $\Omega = -Ng$, where the frequency $|\Omega| = 2\cosh \delta >
2$ lies outside the linear band. Furthermore, $\theta=0$ for the
attractive interactions $g<0$, where the localized solution is the
ground state, whereas $\theta = \pi$ for the repulsive interactions
$g>0$ introduces a staggered phase profile and $\psi_n^{\rm (loc)}$
corresponds to the highest excited state. From the expression $A^2 =
\sqrt{N^2 - 4/g^2}$ for the amplitude, we find that the system
exhibits a threshold for the existence of localized states
\cite{Tsironis94pre}, which are only found for $N>2/|g|$. Since $g$
may be tuned to any small value, the threshold for the number of
particles can be made arbitrarily large. Conversely, for a given
number of particles, there is a threshold value of $g$ for
localization to occur. Figure \ref{fig:FigThresh} shows the
dependence of the energy $E^{(N)}_{\rm class} = \sum_n
-(\psi_{n-1}^{\rm (loc)} \psi_n^{\rm (loc)*} + \text{c.c.}) +
\frac{g}{2}|\psi_0^{\rm (loc)}|^4= \frac{2}{g} +\frac{N^2 g}{2}$ on the
coupling constant in the case of $N=2$ particles. In particular, no
bound state is found classically in this system if $|g|<1$.

\begin{figure}[hbt]
\center
\includegraphics[width=0.49\columnwidth]{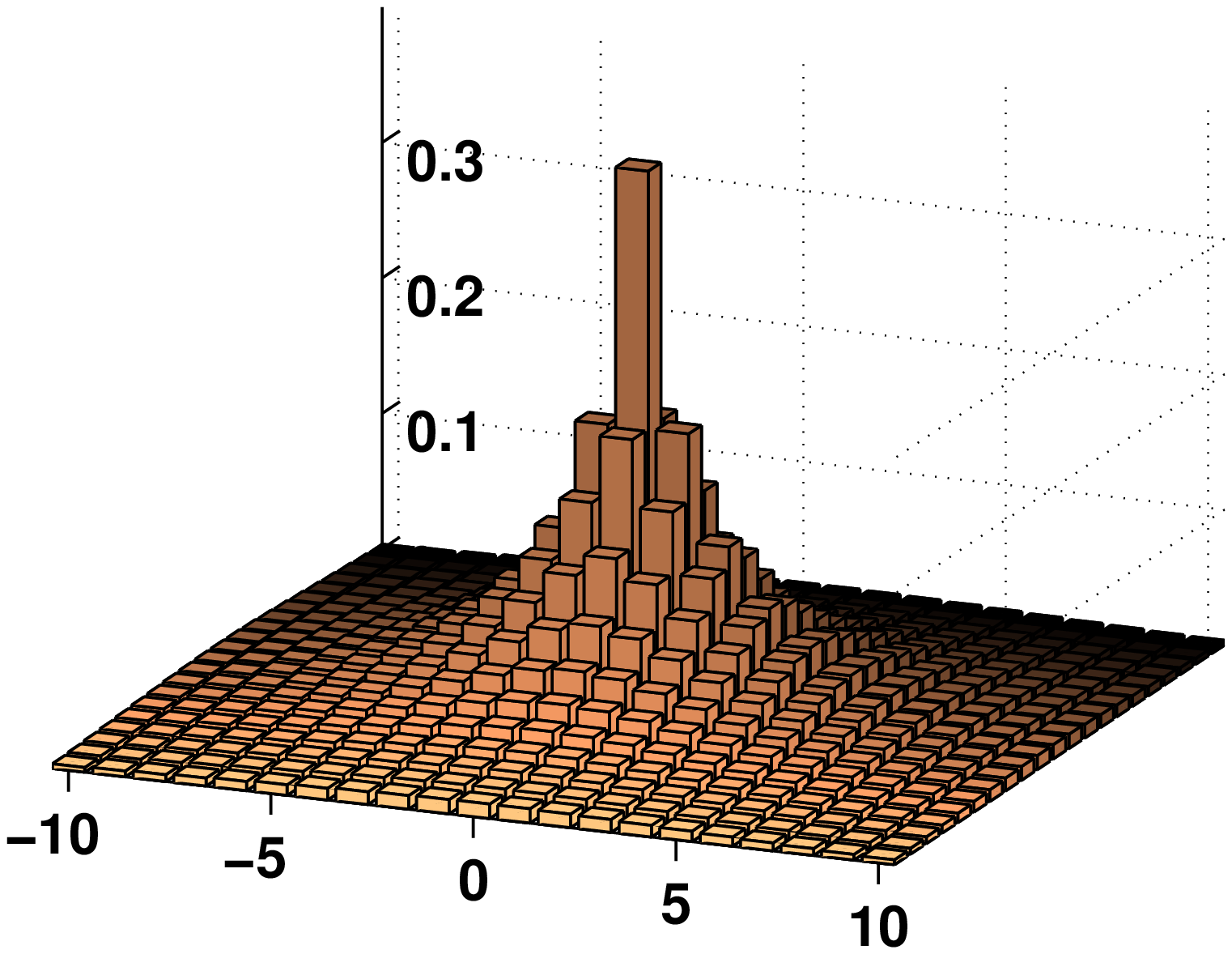}
\includegraphics[width=0.49\columnwidth]{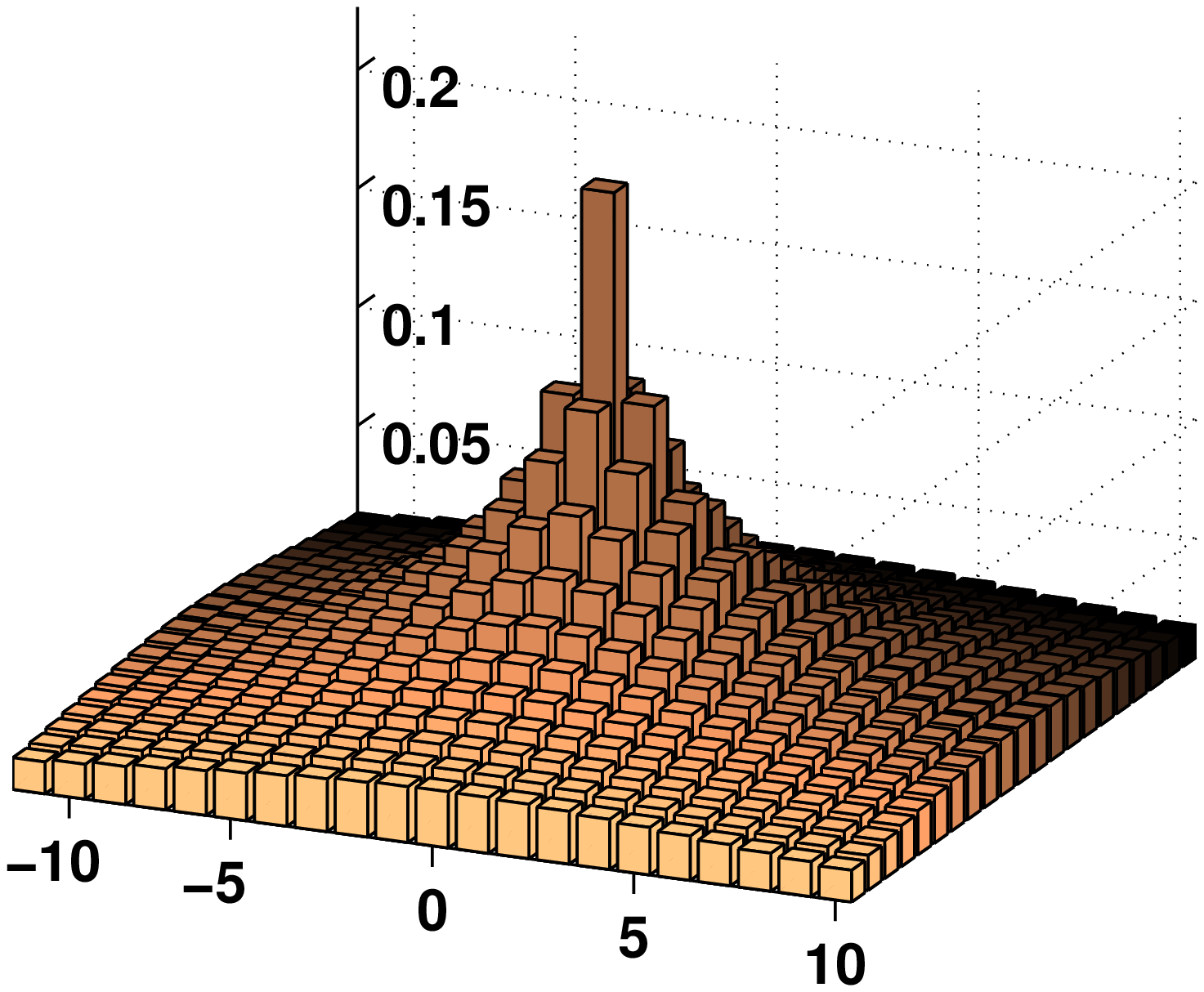}
\\
\includegraphics[width=0.49\columnwidth]{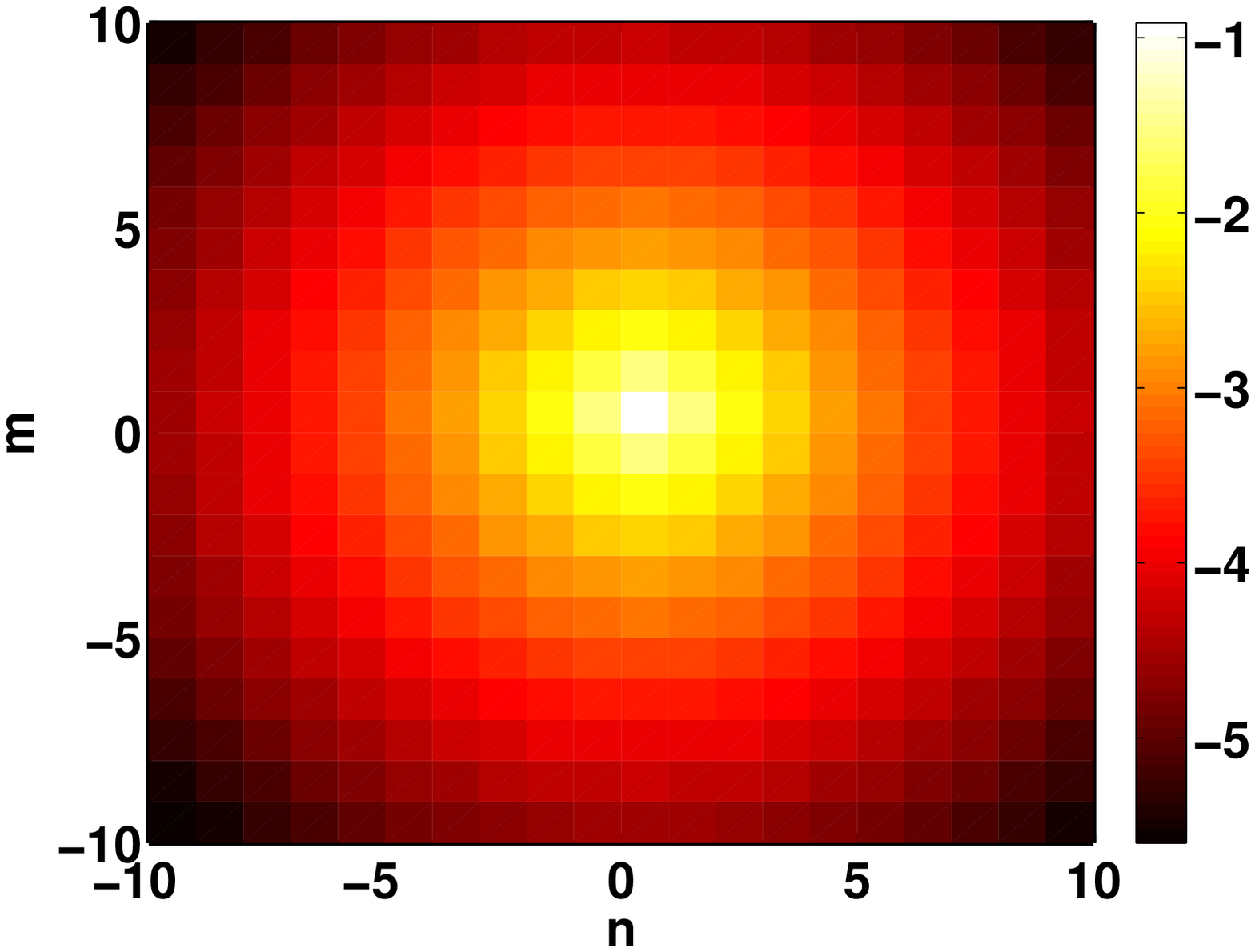}
\includegraphics[width=0.49\columnwidth]{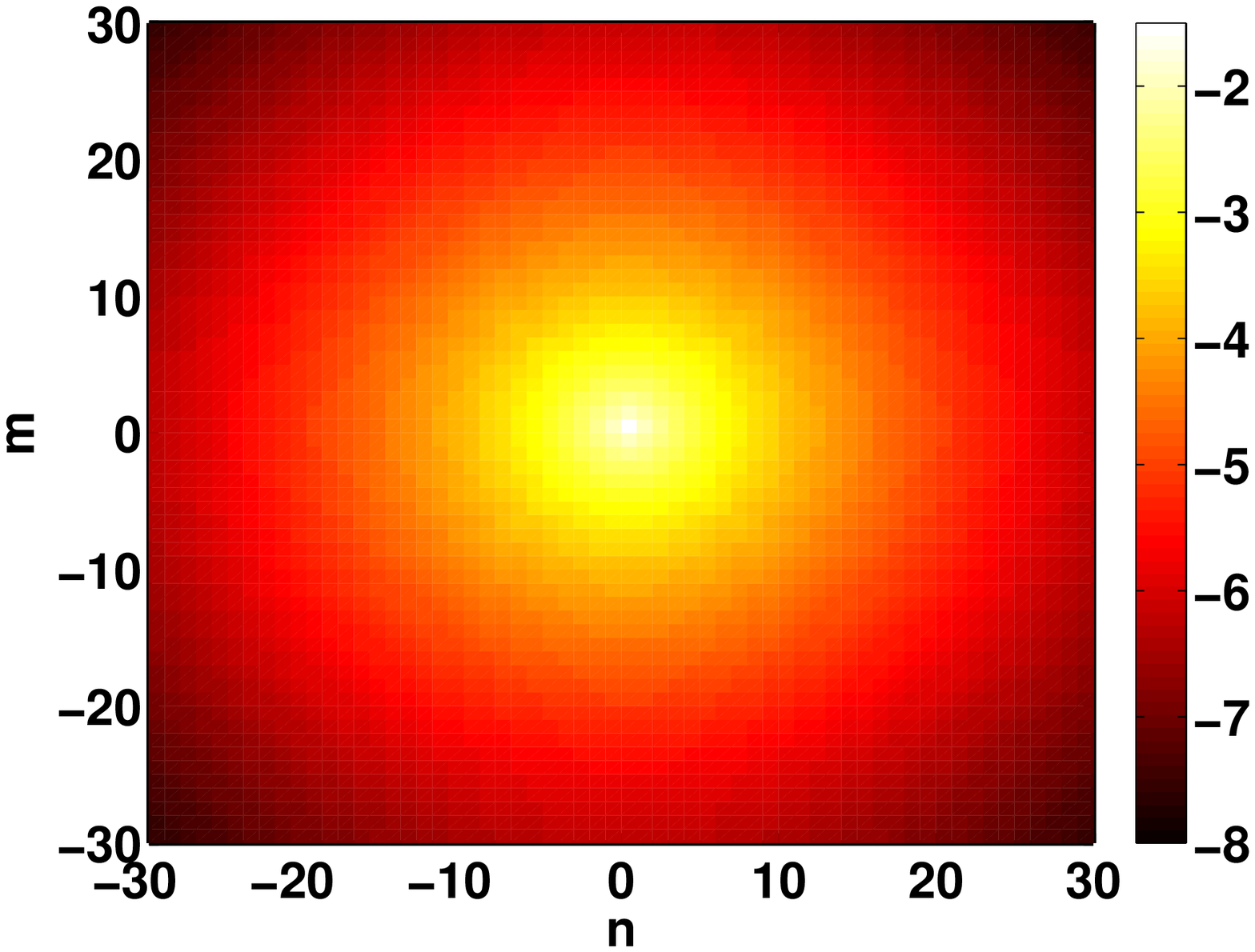}
\caption{Two-particle wave function [top panels:
  $|\varphi_{nm}|$, bottom panels:  $\ln(|\varphi_{nm}|)$]
for the localized state. Left panels: strong localization $\lambda =
-0.978$ and $E=-4.05$. Right panels: weak localization $\lambda =
-0.779$ and $E=-4.01$. The $s$-wave symmetry visible in
the far field is a signature of entanglement, as a product wave
function $\phi_n \phi_m$ would only allow four-fold symmetry. Because
of the (near) cylindrical symmetry of the true wave function, the
kinetic energy cost of bringing additional amplitude to the site
$(0,0)$ is significantly reduced compared to that of the product wave
function.
}
\label{fig:Fig2DWF}
\end{figure}

{\em Quantum case of two particles} - We study the two particle
sector, where we expect to find the most obvious deviations from the
classical theory. In order to solve for the eigenstates $|\Psi^{(2)}
\rangle$ of the Hamiltonian (\ref{eqn:hamiltonian}) for two
particles, we introduce the projected amplitudes or two particle
wave functions $\varphi_{n,m} = \langle{\rm vac}| a_n a_m
|\Psi^{(2)}\rangle$, which obey the equation
\begin{align} \label{eqn:twop}
\nonumber
E \varphi_{n,m} =& -(\varphi_{n,m+1}+\varphi_{n,m-1} + \varphi_{n+1,m} +
\varphi_{n-1,m}) \\ &+ 2 \lambda \delta_{n,0} \delta_{m,0}  \varphi_{0,0}.
\end{align}
This can be interpreted as the Schr\"odinger equation of a single
particle on a two-dimensional lattice with a point defect at the lattice
site $(0,0)$. The problem is known to have a localized solution
for any nonzero value of $\lambda$ \cite{callaway91:QT_solid_state}.
Introducing the Fourier transform
\begin{equation}
\chi_{\mathbf k} = \frac{1}{M} \sum_{m,n} e^{-i\frac{2\pi}{M}(k_1n +
k_2 m)} \varphi_{n,m} ,
\end{equation}
for a square lattice of $M \times M$ sites with ${\mathbf k} =
(k_1,k_2)$ being the quasimomentum vector, eq.~(\ref{eqn:twop})
becomes
\begin{equation}
\chi_{\mathbf k} = \frac{1}{E-{\cal E}_{\mathbf k}} \frac{2
\lambda}{M} \sum_{\mathbf k'}\chi_{\mathbf k'}.
\end{equation}
Looking for localized solutions with $|E|>4$ lying outside the band
of plane-wave energies ${\cal E}_{\mathbf k} = - 2
(\cos\frac{2\pi}{M}k_1 + \cos\frac{2\pi}{M}k_2)$ , we find in the
limit $M\to \infty$
\begin{equation} \label{eqn:Eexact}
\lambda = \frac{1}{2F(E)}\;,\;F(E)=\frac{2}{\pi E} K(16/E^2)
\end{equation}
for the relation of the coupling parameter and the bound-state
energy $E$ (see fig.\ \ref{fig:FigThresh}). Here, $K$ is the complete
elliptic integral of the first kind. It is important to
emphasize that for $4 > |E| > 4.05753$ one has $|\lambda|,|g| <
1$ and thus no classical localized states persist.
However, in the quantum case the asymptotic relation,
\begin{equation}
\label{asymptotics} E(|\lambda| \rightarrow \infty )  \rightarrow 2
\lambda \;,\; |E(|\lambda| \rightarrow 0 )| \rightarrow 4 +
e^{-\frac{2 \pi}{|\lambda|}}\;,
\end{equation}
holds and the localized state wave function is characterized by
\begin{align} \label{eqn:wf}
\varphi_{n,m} = \frac{\sqrt{Z}}{M} \sum_{\mathbf k} \frac{1}{E-{\cal
E}_{\mathbf k}}\, e^{i\frac{2 \pi}{M}(k_1 n + k_2 m)} \,,
\end{align}
with the normalization factor $Z=-1/[M^2 F'(E)]$. The bound-state wave
function $\varphi_{n,m}$ is plotted in fig.~\ref{fig:Fig2DWF} for two
classically forbidden cases.
It is easy to see that this bound and
localized state is the ground state or the highest energy state in the
two-particle sector for attractive ($\lambda <0$) or repulsive
($\lambda>0$) interactions, respectively.
%

{\em Entanglement} - A system of $N$ particles is entangled if the
multi-particle wave function $\phi_{n_1,n_2,\ldots,n_N}$ cannot be
expressed as a product $\phi^1_{n_1} \phi^2_{n_2} \ldots
\phi^N_{n_N}$ of single-particle wave functions. If the state of the
system can be expressed by a product wave function, it is separable.
The Hartree method becomes exact when no entanglement is present.
Since the Hartree method is variational, it gives the {\em best}
separable approximation in the sense that the Hartree energy will be
the closest approximation to the true eigenvalue of the
multi-particle Hamiltonian that can be obtained with a separable
wave function.

\begin{figure}[hbt]
\center
\includegraphics[width=\columnwidth]{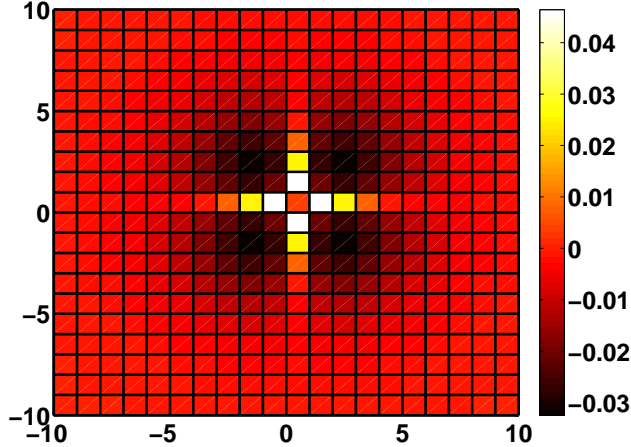}
\caption{{Difference $\varphi_{n,m} - \psi_n^{\rm (loc)}
    \psi_m^{\rm (loc)}$ between the exact two-particle wave function
    and the separable (Hartree) approximation of eqs.~(\ref{eq:psiloc})
    and (\ref{eqn:wf}), respectively, at $\lambda=-1.26$.}}
\label{fig:Diff}
\end{figure}

Let us discuss the two-particle problem. If the two-particle state
were separable, due to Bose symmetry, it would be possible to write it
in the form $\phi_n \phi_m$. This is inconsistent with the result that
in the far field, where the underlying lattice structure becomes less
important, we observe cylindrical ($s$-wave) symmetry as seen in
fig.~\ref{fig:Fig2DWF}. {A separable product approximation,
  on the contrary, is inconsistent with $s$-wave symmetry and is
  characterised by ridge-like structures along the $n=0$ and $m=0$
  co-ordinate axes. These structures are clearly seen in the
  difference between the exact and Hartree two-particle wave functions
  shown in fig.~\ref{fig:Diff}.  }

\begin{figure}[hbt]
\center
\includegraphics[width=\columnwidth]{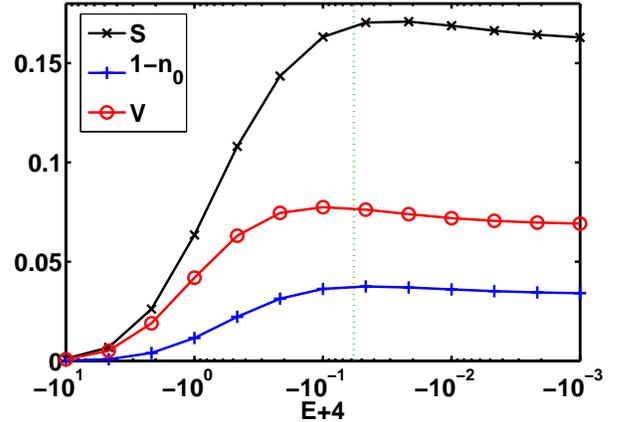}
\caption{Entanglement in the two-particle wave function
  $\varphi_{n,m}$ as a function of the energy $E^{(2)}$.  Above the
  classical threshold $\lambda>-1$ or $E^{(2)}>-4.05753$ (shown as a
  dotted line) entanglement is essential for localization.  Shown are
  the von Neumann entropy $S$, the condensate depletion (geometric
  measure) $1-n_0$, and the entanglement measure $V$ as defined in
  eq.~\ref{eq:defVar}.} \label{fig:Entanglem}
\end{figure}

In order to quantify the entanglement of the two particles in the
bound state (\ref{eqn:wf}) we compute various measures of
entanglement, as shown in fig.~\ref{fig:Entanglem}. First we used the
von Neumann entropy $S = {\rm Tr}(\rho \ln \rho)$
\cite{paskauskas01pra}. Here $\rho$ is the single-particle density
matrix (SPDM) with elements $\rho_{i,j} = F^{-1} \langle
\Psi^{(2)}|a_j^\dagger a_i|\Psi^{(2)}\rangle$, normalized with
$F=\sum_i \langle \Psi^{(2)}|a_i^\dagger a_i|\Psi^{(2)}\rangle$ to
have ${\rm Tr} \rho =1$. Another measure derived from the SPDM is the
condensate depletion $1-n_0$ (also coined geometric measure of entanglement
\cite{Wei03:geomEntanglm}). Here, $n_0$ is the largest eigenvalue of
$\rho$ and measures the fraction of particles in a Bose-condensed
state. Because $\rho$ describes a pure state, $1-n_0$ measures quantum
depletion, which, as we argue here, characterizes quantum
entanglement. This would not be the case in the presence of
incoherent, e.g.\ thermal, excitations.

A third measure, $V$, that is amenable to analytic calculations is
also shown in fig.~\ref{fig:Entanglem}. It uses projected
orbitals defined as $g_n = G^{-1} \sum_m \varphi_{m,n}$, where $G =
\sum_{m,n} \varphi_{m,n}$. Since we expect for separable states that
$\varphi_{m,n}$ is equal to the product $g_n g_m$, where $g_n =
\sum_m \varphi_{nm}$, the deviation
\begin{equation} \label{eq:defVar}
V = \sum_{m,n} |\varphi_{n,m} - g_n g_m|^2
\end{equation}
is a measure of entanglement. Calculating $V$ from eq.~(\ref{eqn:wf})
analytically we find
\begin{equation}
 V \to -7 +\frac{17}{4}\pi - 8 \arctan 1
 \approx 0.06858
\end{equation}
in the limit $\lambda\to 0$, in excellent agreement with the numerical
result shown in fig.\ \ref{fig:Entanglem}. As this Figure 
shows, the different entanglement measures provide a similar picture,
although they are in general not monotonic functions of each other. In
particular we note that the entanglement quickly reaches its maximum
value near the classical threshold. It remains finite as the two
particles become infinitely weakly bound at $\lambda\to 0$.

{\em Three or more particles} - We now show that bound states with
any number of atoms larger than two exist in the quantum model
(\ref{eqn:hamiltonian}) for any value of the coupling constant
$\lambda \neq 0$ as well. Without loss of generality we assume
$\lambda<0$. We have already found a two-body bound state. It will
suffice to show that any bound $p$-particle ground state
$|\psi\rangle$ binds another particle for any $p \ge 2$. For this we
have to find a $(p+1)$ - particle wave function $|\phi\rangle$ with
$\langle\phi|H|\phi\rangle < E^{(p)} - 2$, where $E^{(p)}$ is the
energy eigenvalue of $|\psi\rangle$ and the minimum energy of a free
particle is $-2$. We use the ansatz $|\phi\rangle = \alpha \sum_n
x^{-|n|} a_n^\dagger |\psi\rangle$, which is normalizable if $x>1$.
We choose $\alpha>0$ as a normalization constant to ensure
$\langle\phi|\phi\rangle = 1$. We find that $\langle \phi|H|
\phi\rangle \le E^{(p)} + F(x)$, where $F(x)= (1+ 2 \lambda c)x^2 -
1 -x - x^{-1} + (x - x^{-1})p$ and $c = \langle\phi|a_0^\dagger
a_0|\phi\rangle >0$. Since $F(1)=-2 + 2\lambda c < -2$ it follows
from continuity that there is an $x>1$ such that
$\langle\phi|H|\phi\rangle \le E^{(p)} + F(x) < E^{(p)} - 2$ as
required. This concludes the proof that bound states with any
particle number exist in the quantum problem.

We further remark that having found an $N$-particle bound state that
persists below the classical threshold $\lambda<\lambda_{\rm thresh} =
1/(N-1)$, we 
automatically know that entanglement plays an essential role in its
binding. This is because the best separable wave function is in fact
the Hartree approximation, which does not bind there.

{ {\em Translationally invariant systems} - In interacting
  lattice problems with translational invariance the quantum
  eigenstates are delocalized due to fundamental properties of quantum
  theory. However, the existence of lattice solitons in a
  corresponding classical theory indicates the existence of quantum
  states with local second order correlation known as quantum lattice
  solitons, which can be interpreted as bound states of quantum
  particles \cite{scott94:_quant,lai89praI,lai89praII}.  Both lattice
  solitons and quantum lattice solitons are characterised by
  frequencies and energies, respectively, outside of the bands of
  delocalized solutions in the noninteracting system. A framework for
  detailed comparison between the thresholds predicted by classical
  and quantum theory is, again, enabled by establishing the classical
  theory as a Hartree approximation to the quantum problem. In the
  classical theory, there is no threshold in a one dimensional lattice
  with a cubic nonlinearity (corresponding to two-particle
  interactions) but there are thresholds for higher dimensions
  \cite{flach97:thresholds}.  The variational properties of the
  Hartree approximation guarantee that the existence of lattice
  solitons in the classical theory implies the existence of quantum
  lattice solitons but not vice versa. If quantum solitons exist below
  a classical threshold in these systems we thus know that
  entanglement between quantum particles plays a vital role. However,
  we also expect entanglement to be relevant for delocalised quantum
  soliton states above the thresholds (i.e. for stronger
  interactions). It is known that thresholds for quantum solitons
  exist in dimension higher than one
  \cite{hecker-denschlag07:Varenna}.
   
  Extending the current model with spatially localized interactions
  into more than one dimensions, there will generally be thresholds
  for localization in both the quantum and the classical
  models\cite{dorignac08phd}.  However, these thresholds will
  generally differ.  The detailed study of such systems lies beyond
  the scope of this letter and presents an interesting opportunity for
  future work.  }

In {\em conclusion}, we have shown that localized states of a few
atoms in an optical lattice with spatially confined $s$-wave
interaction persist below the classical threshold. Moreover, wave
function entanglement plays a crucial role in that localization.
A one-dimensional optical lattice with spatially inhomogeneous
interactions can be engineered with presently available techniques
using magnetic or optically-induced Feshbach resonances
\cite{morsch06rmp}.
Increasing the size of the spatial interaction domain will decrease
the classical threshold, but it will stay finite. Thus quantum
localization by entanglement is robust and will disappear only in
the limit of an infinite interaction domain, where the classical
model is known to have zero thresholds for localized states
\cite{flach97:thresholds}.
{In an
experiment where interactions are tuned below the classical threshold
the observation of localized modes
will indicate the vital role of entanglement.
This} entanglement between atoms is
distillable \cite{paskauskas01pra} and could possibly be measured
with entanglement witnesses or by reconstruction of the
single-particle density matrix from position and momentum-space
measurements. Beyond the currently studied model we expect that
quantum entanglement favors localization in other quantum lattice or
quantum field theories as well.

{ This work was partially supported by the Marsden Fund of
  New Zealand under contract number MAU0706 and by NSF Grant PHY 05
  55313. V.F.\ was supported by the Israeli Science Foundation, grant
  No.\ 0900017.}

\bibliographystyle{eplbib}

\end{document}